# Phosphorene as a superior gas sensor: Selective adsorption and distinct I-V response

Liangzhi Kou,*,† Thomas Frauenheim,† and Changfeng Chen‡

*Bremen Center for Computational Materials Science, University of Bremen, Am Falturm 1, 28359 Bremen, Germany*

*Department of Physics and Astronomy and High Pressure Science and Engineering Center, University of Nevada, Las Vegas, Nevada 89154, United States*

E-mail: kouliangzhi@gmail.com

Recent reports on the fabrication of phosphorene, i.e., mono- or few-layer black phosphorus, have raised exciting prospects of an outstanding two-dimensional (2D) material that exhibits excellent properties for nanodevice applications. Here we study by first-principles calculations the adsorption of CO, $CO_2$, $NH_3$, NO and $NO_2$ gas molecules on a mono-layer phosphorene. Our results predict superior sensing performance of phosphorene that rivals or even surpasses other 2D materials such as graphene and $MoS_2$. We determine the optimal adsorption positions of these molecules on the phosphorene and identify molecular doping, i.e., charge transfer between the molecules and phosphorene, as the driving mechanism for the high adsorption strength. We further calculated the current-voltage (I-V) relation using a non-equilibrium Green's function (NEGF) formalism. The transport features show large (one to two orders of magnitude) anisotropy along different (armchair or zigzag) directions, which is consistent with the anisotropic electronic band structure of phosphorene. Remarkably, the I-V relation exhibits distinct responses with a marked change of the I-V relation along either the armchair or the zigzag directions depending on the type of molecules. Such selectivity and sensitivity to adsorption makes phosphorene a superior gas sensor that promises wide-ranging applications.

KEYWORDS: Phosphorene, Gas Sensor, Anisotropic Transport

# Introduction

Two-dimensional (2D) materials hold great promise in future nanodevice applications due to their high mobility, outstanding mechanical performance and large surface-to-volume ratio,[1] as demonstrated in extensive studies of graphene and transition-metal dichalcogenides ($MoS_2$, $WSe_2$ and so on).[2–4] The utility of these materials, however, are limited by some intrinsic shortcomings, such as the lack of a bandgap in graphene[2] and the relative low mobility in $MoS_2$,[5] which has motivated continuing work in search of more 2D materials that exhibit properties that may lead to specific improved performance. Recent experiments reported successful fabrication of few-layer black phosphorus (also known as phosphorene).[6–8] Black phosphorus shares a structural feature with graphite in that it is also a layered material with weak interlayer van der Waals (vdW) interaction, which allows the fabrication of phosphorene by micromechanical cleavage and exfoliation methods. In monolayer phosphorene, each phosphorus atom forms bonds with three adjacent phosphorus atoms in a puckered honeycomb structure. Phosphorene, however, has a significant advantage over the semimetallic graphene since it exhibits a finite and direct band gap within an appealing energy range[9–11] and its measured free-carrier mobility (around 1000 cm$^2$/v.s)[6] is better than other typical 2D semiconductors, such as MoS$_2$ (around 200 cm$^2$/v.s).[5] Phosphorene also exhibits other interesting and useful features, including its anisotropic electric conductance and optical responses,[6,7,11] which distinguishes it from other isotropic 2D materials such as graphene and molybdenum and tungsten chalcogenides. These excellent properties have already been exploited to find important applications in field effect transistor (FET)[6] and thin-film solar cells.[12]

2D materials are also usually good candidates for gas sensors due to their large surface-to-volume ratio and the associated charge transfer between gas molecules and the substrates. For graphene[13,14] and MoS$_2$,[15–17] good sensor properties have already been demonstrated by both theoretical and experimental investigations. In these examples, it was shown that the charge carrier concentration induced by gas molecule adsorption can be used to make highly sensitive sensors, even with the possibility of detecting an individual molecule, where the sensor property is based on changes in the resistivity with the gas molecules acting as donors or acceptors. It is expected that the electrical resistivity of phosphorene will also be influenced by the gas molecule adsorption in a similar way. Given the distinctive electronic properties of phosphorene, it is highly desirable to explore and establish the trends and rules of gas molecule adsorption on phosphorene and distinct characteristics of the influence of the molecules on the transport behavior, which can be used as the markers in sensing applications.

In this Letter, we report first-principles calculations that examine the adsorption of several typical molecules, CO, $CO_2$, $NH_3$, NO and $NO_2$, on phosphorene. We first determine their preferential binding positions and the corresponding binding energy. Our results show that the strength of binding is highly dependent on the amount of charge transfer between the molecules and the phosphorene layer. This is similar to the situation observed in graphene and $MoS_2$, but the adsorption on phosphorene is generally stronger, which tends to have a more pronounced influence on the property of the host layer (i.e., phosphorene in the present case), making it a more sensitive sensor. We calculated the I-V relation of phosphorene without and with the gas molecule adsorption using a non-equilibrium Green's function (NEGF) formalism. The results not only show sensitive changes to adsorption, but remarkably also exhibit high selectivity in that different gas molecules may induce changes in transport behavior along either the armchair or zigzag direction of the phosphorene layer. The combined sensitivity and selectivity of the I-V relation to adsorption of various gas molecules make phosphorene a promising candidate for high-performance sensing applications.

## Methods

Structural relaxation and electronic calculations were carried out by first-principles calculations based on the density functional theory (DFT) as implemented in the Vienna Ab Initio Simulation (VASP) package.[18] The exchange correlation interaction was treated within the local-density approximation (LDA). The structural model for monolayer phosphorene is periodic in the *xy* plane and separated by at least 10 Å along the *z* direction to avoid the interactions between adjacent layers. All the atoms in the unit cell are fully relaxed until the force on each atom is less than 0.01 eV/Å. The Brillouin-zone integration was sampled by a 10 × 8 × 1 *k*-grid mesh. An energy cutoff of 400 eV was chosen for the plane wave basis. The vdW interaction is introduced to treat the interaction between the gas molecules and the phosphorene layer, and it is described by a semiempirical correction by the Grimme method.[19] Spin polarization was included in the calculations of the adsorption of NO and $NO_2$ since these molecules are paramagnetic, but not considered in the calculations for other gas molecules. The electronic transport properties are studied by the non-equilibrium Green's function (NEGF) techniques within the Keldysh formalism as implemented in the TRANSIESTA package.[20,21] The electric current through the contact region is calculated using the Landauer-Buttiker formula[22]

$$I(Vb) = G0 \int_{\mu L}^{\mu R} T(E, Vb) dE$$

where $G_0$ is the unit of the quantum conductance and $T(E, V_b)$ is the transmission probability of electrons incident at energy E under a potential bias $V_b$. The electrochemical potential difference between the two electrodes is $eV_b = \mu_L - \mu_R$.

## Results and discussion

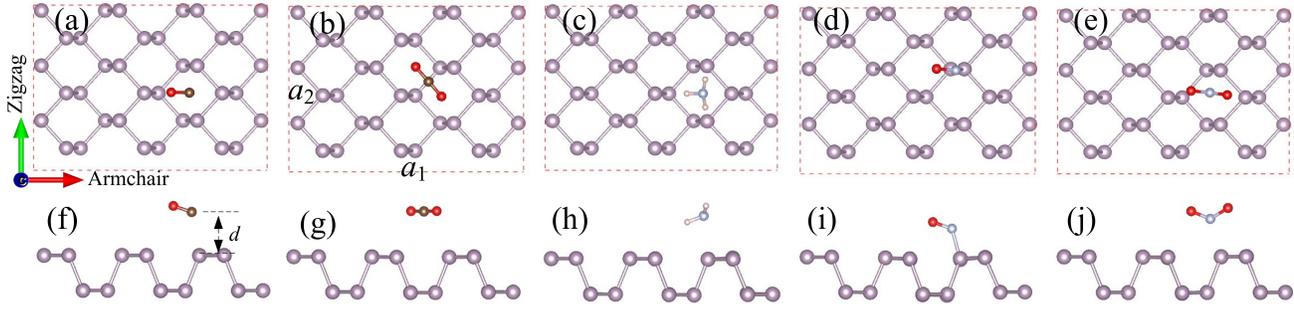

Figure 1: Top view (a-e) and side view (f-j) of the fully relaxed structural models of phosphorene with CO, $CO_2$, $NH_3$, NO and $NO_2$ adsorption, respectively. The red dashed rectangle is the supercell in the present work. The brown balls represent P atoms while the black, red, cyan and light white balls indicate C, O, N and H atoms, respectively. The distance between the gas molecule and the phosphorene layer is indicated in (f).

To model the phosphorene layer, we used a 3 × 3 surpercell as shown in Fig. 1, which has lattice dimensions of 9.898 Å × 13.854 Å corresponding to the unit cell of 3.3 Å × 4.618 Å, which is in good agreement with previous reports.[9,10] The boundary structure is anisotropic along the $x$ and $y$ directions, labeled as armchair and zigzag direction, respectively. For each adsorption case, a molecule is placed near the phosphorene layer, and the whole system is fully relaxed. We present all the relaxed structures with different gas molecule adsorption in Fig. 1. For CO molecule, the carbon atom is located at the center of the puckered honeycomb. But the carbon atom of $CO_2$ is located at the bridge of the P-P bonding. The distance between the CO molecule and the phosphorene layer (2.56 Å) is slightly smaller than the value for $CO_2$ adsorption (2.69 Å). In contrast, for another dipolar molecule NO, it moves to the top position of a P atom, forming a direct bond with it, see Fig. 1(d) and 1(i). As a result, the adsorption distance is considerably reduced to 1.73 Å, which is in the range of the P-N bonding distance. However, for $NO_2$, it does not bond with the P atom due to the lack of dangling bonds and, consequently, stays above the phosphorene layer with an adsorption distance of 2.2 Å, as shown in Fig. 1(e) and 1(j). The $NH_3$ molecule is also staying above the phosphorene layer with an adsorption distance of 2.14 Å with the N atom located at the center of the puckered honeycomb, as seen in Fig. 1(c) and 1(h).

For a quantitative description of the adsorption strength on phosphorene and a comparison with adsorption on other 2D materials, such as graphene and $MoS_2$,[13–17] we calculated the adsorption energy ($E_a$), which is defined as the energy of the isolated phosphorene layer ($E_P$) and the isolated molecule ($E_{gas}$) minus the energy of the fully relaxed phosphorene with the gas molecule adsorption ($E_{P+gas}$), $E_a = E_{gas} + E_P - E_{P+gas}$. The obtained results are displayed in Fig. 2(a). One can see that CO has the weakest adsorption strength of 0.325 eV/unit cell while NO has the largest adsorption energy of 0.863 eV/unit cell among the gas molecules studied in the present work. The values for $CO_2$, $NH_3$ and $NO_2$ adsorption are 0.41, 0.5 and 0.62 eV/unit cell, respectively. The N-based gases generally have larger adsorption energies than those for CO and $CO_2$ adsorption, indicating that phosphorene is more sensitive to the nitrogen-based toxic gases, which is similar to $MoS_2$ that is extremely sensitive to adsorption of $NO_x$ and $NH_3$ and can detect them down to the concentration of 1 ppb.[16,17] Interestingly, we notice that the adsorption energies on phosphorene are generally larger than those on other 2D materials,[13,23] suggesting a higher level of sensitivity for gas adsorption on phosphorene. Previous studies of gas adsorption on graphene and $MoS_2$ identified the important role of charge transfer in determining the adsorption energy and causing a decrease in the resistance of $MoS_2$. Here we examine this issue for adsorption on phosphorene. In Fig. 2 (b) to 2(f), we show the calculated charge transfer, which is defined as $\Delta\rho = \rho_{tot}(r) - \rho_{BP}(r) - \rho_{gas}(r)$, for the adsorption of CO, $CO_2$, $NH_3$, NO and $NO_2$, respectively. A comparison of these results with that shown in Fig. 2(a) shows a clear correlation between the adsorption strength (energy) and the amount of charge transfer. For CO adsorption on phosphorene, there is only a small amount of charge transfer (0.03 $e$) from the CO molecule to the phosphorene layers, causing a weak binding. The electron transfer for $CO_2$ adsorption increases slightly to 0.04 $e$, producing a corresponding enhancement in the binding as reflected in a stronger adsorption energy. When we look into the nitrogen based gas molecules, a much more significant charge transfer is observed. Especially for NO, which has the strongest adsorption energy (Fig. 2e), up to 1 $e$ is transferred from the phosphorene layer to the molecule. For the relatively weaker adsorption of $NH_3$ and $NO_2$, the electron transfer is correspondingly smaller than that for NO, but still larger than those for CO and $CO_2$, as shown in Figs. 2(d) and 2(f). This systematic trend in adsorption strength correlated with the charge transfer expands our understanding of the mechanism for gas molecule adsorption on phosphorene; it also provides an avenue for electric field control of gas adsorption as demonstrated in $MoS_2$.[24]

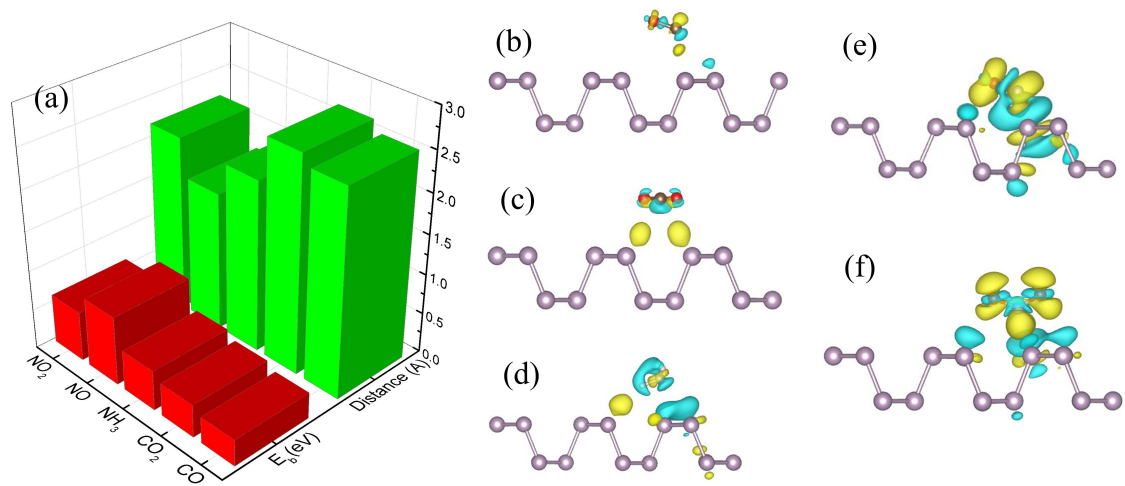

Figure 2: (a) Distance between the gas molecule and the phosphorene layer as defined in Fig. 1(f) and the adsorption energy for CO, $CO_2$, $NH_3$, NO and $NO_2$ on phosphorene. The adsorption configurations and charge transfer for each case are plotted in (b) to (f). The isosurface value for all the cases is $10^{-3}$ $e$/Å$^3$. The yellow isosurface indicates an electron gain while blue one represents an electron loss.

We now turn to the effects of gas adsorption on the electronic properties of phosphorene. We show in Fig. 3(a-c) the total density of states (DOS) of the phosphorene monolayer with the adsorption of the non-paramagnetic CO, $CO_2$ and $NH_3$ molecules, as well as the projected DOS from various gas molecules. The results show a bandgap of 1.0 eV, in accordance with previous results of pure phosphorene.[9–11] The DOS for either the valence or conduction bands of phosphorene is not significantly influenced upon the CO or $CO_2$ adsorption, which is consistent with their small adsorption energies. Meanwhile, the adsorption of $NH_3$ molecule induces several distinct states at the lower-lying valence bands in the energy range around -8 eV. The adsorptions of these three molecules produce no noticeable modifications of the DOS near the Fermi level. As a result, it can be concluded that the adsorption of CO, $CO_2$ and $NH_3$ does not have a substantial effect on the electronic structure of phosphorene.

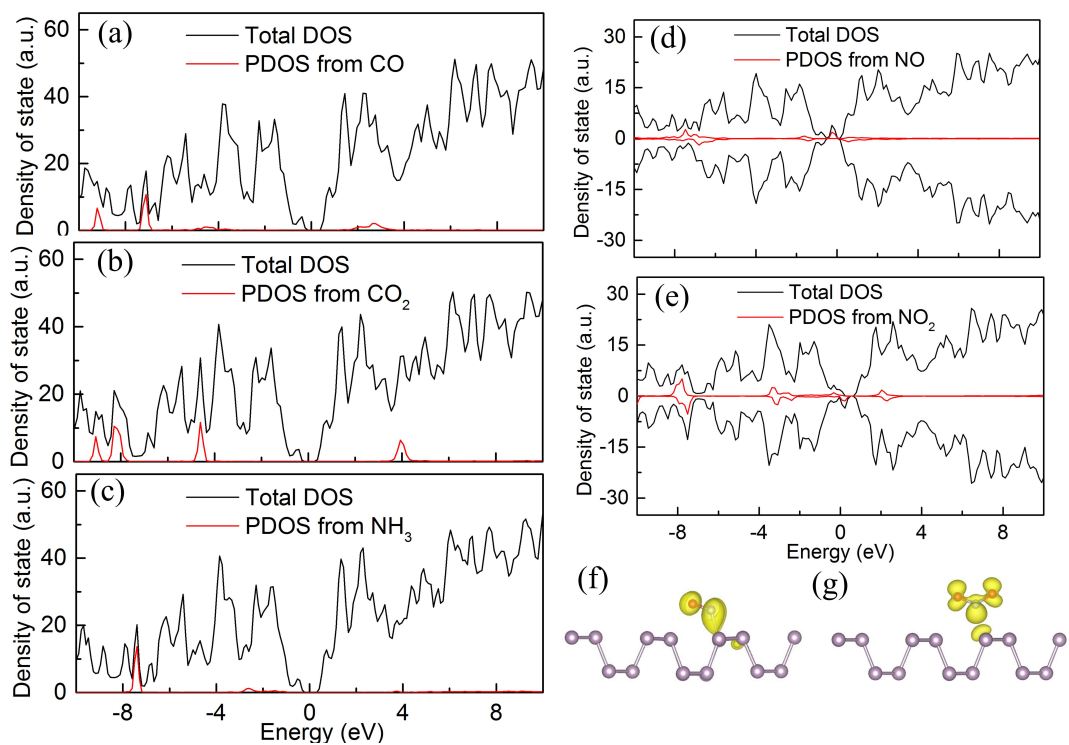

Figure 3: (a-e) Total density of states (DOS) of phosphorene with each gas molecule adsorption (black lines) and the projected DOS from the gas molecule (red lines). (f,g) The spatial spin density distribution on NO and $NO_2$ molecules.

On the other hand, the adsorption of paramagnetic molecules NO and $NO_2$ leads to a higher level of doping of the phosphorene layer, which is manifested in their higher adsorption energy. The spin-polarized DOS plots of NO and $NO_2$ adsorption on phosphorene are displayed in Figs. 3(d) and 3(e), with the distribution of spin density included in Figs. 3(f) and 3(g), respectively. It is seen that the magnetization is mainly located on the NO or $NO_2$ molecules in both cases. For the NO molecule on phosphorene, the total magnetic moment is 1 $\mu B$. A spin-up impurity state occurs near the Fermi level, which can be identified in the projected DOS plot. The differences around the Fermi level are mainly attributed to the N-p orbitals according to a projected DOS analysis, which results in a DOS peak for the spin-up electrons. Similarly the adsorption of $NO_2$ on phosphorene also leads to a magnetic moment of 1 $\mu B$, and the Fermi level is downshifted to the conduction band, indicating an electron doping for the phosphorene layer.

Although the electronic properties of phosphorene after the adsorption of CO, $CO_2$ and $NH_3$ are not substantially changed, the adsorption induced charge transfer is expected to affect the resistivity of the system, which can be measured experimentally and act as a marker for gas sensors. To explicitly evaluate the performance of phosphorene as a gas sensor, we employed the NEGF method to calculate the transport transmission and the corresponding current-voltage (I-V) relation before and after the gas adsorption, which allows the monitoring of the resistivity change. The obtained results can be directly compared to experimental measurements. Two representative cases of gas adsorption on phosphorene ($NH_3$, non-spin-polarized; and NO, spin-polarized) are chosen for transport calculations. Due to the structural anisotropy of phosphorene, two transport models are constructed in each case: one has the current flowing along the armchair direction, as shown Fig. 4(a), and the other has the current flowing along the zigzag direction, as shown in Fig. 4(b). For both cases, we use a two-probe system where semi-infinite left and right electrode regions are in contact with the central scattering region, and a 3 × 3 surpercell without gas adsorption is used for each of the left and right electrode, while the center scattering is considered in a 3 × 3 surpercell with gas adsorption. The supercell configuration here is the same as that obtained from the structural relaxation and that used in the electronic calculations. For comparison, calculations were also performed for a 3 × 3 center scattering region without gas adsorption. In Fig. 4(c), we present the I-V curve along the armchair direction of phosphorene with and without the $NH_3$ adsorption. When a bias voltage is applied, the Fermi level of the left electrode shifts upward with respect to that of the right electrode. Therefore, the current starts to flow only after the valance band maximum (VBM) of the left electrode reaches the conduction band minimum (CBM) of the right electrode. As a result, there is no current passing through the center scattering region when the bias voltage is smaller than about 1.0 V, which is the value of the band gap of phosphorene. As the bias voltage further increases, the currents in both spin channels increase quickly. Under a bias of 2.1 V, the current passing through the pure phosphorene is 5.26 $\mu A$; but when with the adsorption of $NH_3$ molecule, the current under the same bias is reduced to 4.7 $\mu A$, which is about 11% reduction. The reduction of current indicates the increase of resistance after the $NH_3$ adsorption, which can be directly measured in experiment. To understand this result, we consider the zero-bias transmission spectra of phosphorene with and withour the $NH_3$ adsorption. Fig. 4(d) shows the transmission spectra of pure phosphorene (blakc line) and phosphorene wit the $NH_3$ adsorption (red line) under zero bias. It is seen that there is a region of zero transmission around the Fermi level with a width of 1.0 V, and beyond this region, there are steplike characteristics in the transmission spectra, which are produced by the available conductance channels of various energy bands. It is clear that some of the conductance channels are partially inhibited by the $NH_3$ adsorption, and it is especially effective at the conduction band region of around -2 eV. The reduced conductance channels lead to the reduction of passing current. In other words, the $NH_3$ adsorption introduces back-scattering centers that reduce available conductance channels, yielding reduced current.

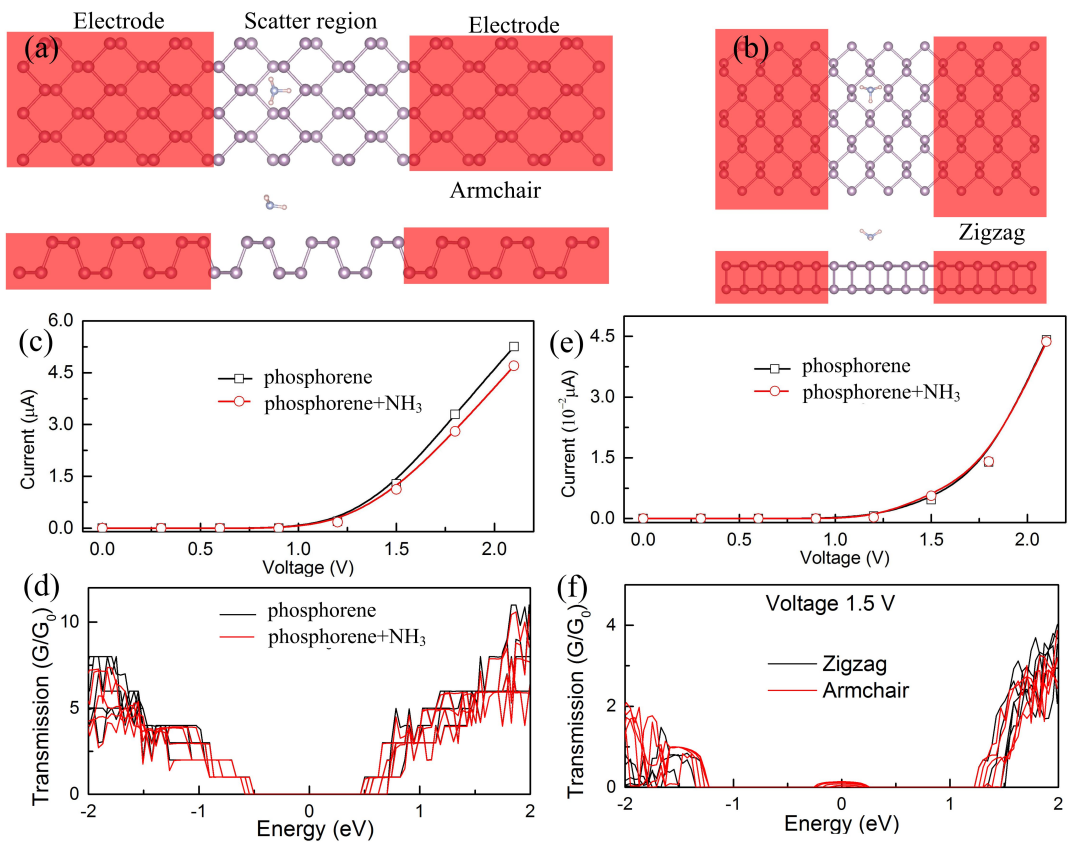

Figure 4: (a,b) Illustration of the two-probe systems where semi-infinite left and right electrode regions (red shade region) are in contact with the central scattering region along the armchair or zigzag directions. For the electrodes and scatter regions, 3 × 3 surpercells of phosphorene are used. The upper panels are top views while lower ones are side views. In (c) and (e), we display the I-V curves along the two directions of pure phosphorene and phosphorene with the $NH_3$ adsorption. The transmission spectra under zero bias are shown

in (d). To understand the anisotropic I-V curves, the transmission spectra along the armchair and zigzag directions are presented in (f).

For the transport along the zigzag direction, an I-V curve similar in functional dependence as that along the armchair direction is obtained [Fig. 4(e)]. The current stays nearly zero under a bias of less than 1.0 V, and it increases quickly under further rising bias. There is, however, a large difference in the size of the current along the zigzag direction, which is about two orders of magnitude smaller than the corresponding values along the armchair direction. For instance, the current of phosphorene with the $NH_3$ adsorption along the zigzag direction under the bias of 2.1 V is $4.37 \times 10^{-2} \mu A$, while the corresponding value along the armchair direction is 4.7 $\mu A$. Interestingly, the $NH_3$ adsorption only induces a minimal current reduction ($4.41 \times 10^{-2} \mu A$ for pure phosphorene along the zigzag direction under the bias of 2.1 V). The anisotropic transport properties of phosphorene originate from its anisotropic electronic band structures (see Fig. S1 in the Supporting Materials), which are highly anisotropic: both the top of the valence bands and the bottom of the conduction bands have much more significant dispersions along the $\Gamma$-Y direction, which is the armchair direction in the real space as indicated in Fig. 1(a); meanwhile, these bands are nearly flat along the $\Gamma$-X direction, which is the zigzag direction in the real space. Therefore, the corresponding effective mass of electrons and holes is also highly anisotropic because it is proportional to the inverse of the curvature of the band dispersion. This directly leads to the anisotropic I-V curve and the associated resistance, which has been confirmed by recent experimental measurements.[7] To better understand the anisotropic transport behavior, we plotted the transmission spectra of the phosphorene with the $NH_3$ adsorption along the zigzag and armchair directions under the bias of 1.5 V, as shown in Fig. 4(f). Along the armchair direction, the non-zero transmission spectra emerges around the Fermi level while the spectra are larger in the deep valance region around -2 eV than those along the zigzag direction. The much larger conductance channels naturally lead to much larger current.

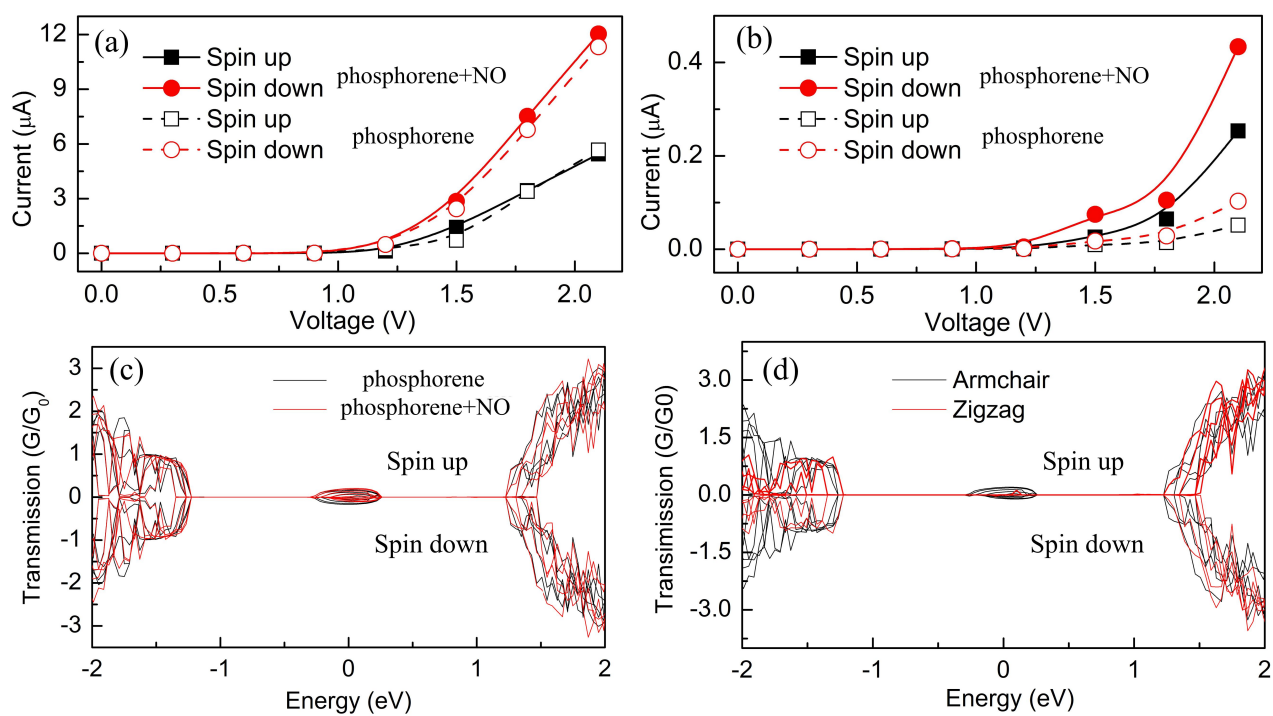

Figure 5: Spin-polarized I-V curves of phosphorene without (dashed lines) and with (solid lines) the NO adsorption along the armchair (a) and the zigzag (b) directions. The transmission spectra along the armchair direction under the bias of 1.5 V are displayed in (c), while a comparison between transmissions along the armchair and the zigzag directions under the bias of 1.5 V is presented in (d).

The adsorption of paramagnetic molecules NO and $NO_2$ on phosphorene induce spin polarization as indicated above, which leads to spin-polarized current. We take phosphorene with the NO adsorption as a representative example to explore the spin-dependent anisotropic transport and I-V curve. Here we also use two models, one along the armchair and the other along the zigzag direction as discussed above. We calculated the I-V curves for spin-polarized phosphorene with and without the NO adsorption for comparison. The spin-polarized I-V curves along the armchair direction are presented in Fig. 5(a). Regardless of the NO adsorption, the currents are zero when the bias is smaller than 1.0 V, the same as in the case of $NH_3$, which is set by the intrinsic gap of phosphorene. As the bias voltage increases, the spin-dependent currents occur, and the values of pure phosphorene under the bias of 1.5 V are 0.71 and 2.44 $\mu A$ for spin-up and spin-down currents, respectively. With the adsorption of NO, the corresponding currents increase to 1.42 and 2.85 $\mu A$ for the two spin directions. This adsorption induced current increase is in sharp contrast to the results for phosphorene with the adsorption of $NH_3$ where there is a considerable current decrease along the armchair direction and essentially no change in the zigzag direction. The current increase induced by the NO adsorption can be attributed to the introduction of spin states inside the gap, as indicated by the projected DOS analysis in Fig. 3(d). These additional states increase the conductance channels for electron transport. From the corresponding transmission spectra in Fig. 5(c), one can also see that the transmission value of phosphorene after the NO adsorption is larger than those of pure phosphorene under the bias of 1.5 V. We now examine the current along the zigzag directions, and the spin-polarized currents increase significantly after the NO adsorption. The currents of $9.68 \times 10^{-3} \mu A$

(spin up) and 1.76 ×10$^{-2}$μ$A$ (spin down) of pure phosphorene under the bias of 1.5 V increase to 2.59 ×10$^{-2}$μ$A$ and 7.47 ×10$^{-2}$μ$A$, respectively, after the NO adsorption, as shown in Fig. 5(b). Meanwhile, there is also a remarkable anisotropy in the I-V curve (namely the anisotropic transport) along the armchair and zigzag directions, which is clearly seen by comparing Fig. 5(a) with Fig. 5(b). The current in the two directions differ by more than ten times. The transmission spectra for the two directions under the same bias are shown in in Fig. 5(d), and larger conductance channels are observed, especially around the energy of -2 eV for currents along the armchair direction with the NO adsorption.

In summary, using first-principle calculations, we have systematically studied the structure, electronic and transport properties of monolayer phosphorene with the adsorption of gas molecules CO, CO$_2$, NH$_3$, NO, and NO$_2$. Our results show that the binding of nitrogen-based gas molecules, such as NO and NO$_2$, is the strongest among the gas molecules considered, suggesting that phosphorene is more sensitive to these toxic gases. This behavior is attributed to the more sensitive change in the electronic band structure and charge transfer induced by the adsorption of these gas molecules. Transport calculations indicate that the gas molecule adsorption on phosphorene can either reduce (e.g., by NH$_3$) or increase (e.g., by NO) the current, thus changing the resistance, which can be directly measured experimentally. The current and the adsorption induced current change of phosphorene exhibit strong anisotropic characteristics along the armchair or zigzag directions of the phosphorene layer. Such sensitivity and selectivity to gas molecule adsorption make phosphorene a desirable candidate as a superior gas sensor.

## Acknowledgement

Computation was carried out at HLRN Berlin/Hannover (Germany). L.K. acknowledges financial support by the Alexander von Humboldt Foundation of Germany. C.F.C. was partially supported by the DOE under the Cooperative Agreement DE-NA0001982.

## Supporting Information Available

The electronic band structure of the unit cell phosphorene is presented, where the anisotropy along the armchair and zigzag directions is clearly seen.

This material is available free of charge via the Internet at http://pubs.acs.org/.